\documentclass{jps-cp}
\usepackage{txfonts} %Please comment out this line unless the txfonts package is availabe in your LaTeX system.
\usepackage{bm}
\usepackage{amsmath}
\allowdisplaybreaks[4]

\title{Exploring the Quark Transversity and the Collins Fragmentation Functions using Polarized $pp$ Collisions at STAR}

\author{Ting Lin for the STAR Collaboration}

\inst{Institute of Frontier and Interdisciplinary Science \& Key Laboratory of Particle Physics and Particle Irradiation (MoE), Shandong University, Qingdao, Shandong 266237, China}

\email{tinglin@sdu.edu.cn}

\recdate{January 15, 2022}

\abst{Understanding the internal spin structure of the nucleon still remains a challenge in strong interaction physics. Transversity, which describes the transverse spin structure of quarks in a transversely polarized proton, is poorly constrained by experimental data. Since it is chiral-odd, it can only be accessed through channels that couple to other chiral-odd distributions, like the Collins fragmentation functions (so-called Collins effect) or the interference fragmentation functions. Recently, a detailed calculation using the soft-collinear effective theory found that the Collins effect in $pp$ collisions involves a mixture of collinear and transverse momentum dependent (TMD) factorization. The Collins effect provides a direct probe to the Collins fragmentation function and enables testing of its evolution, universality and factorization breaking in the transverse momentum dependent formalism. In 2018, STAR published the first measurements of Collins asymmetries for charged pions in jets in polarized $pp$ collisions at $\sqrt{s}$ = 500 GeV based on data taken during 2011. These measurements probe $Q^2$ scales one to two orders of magnitude larger than similar measurements in semi-inclusive deep-inelastic scattering (SIDIS) and the results are consistent with predictions based on global analyses of $e^{+}e^{-}$ and SIDIS data. In 2012 and 2015, STAR collected $\sim$14 $\mathrm{pb^{-1}}$ and $\sim$52 $\mathrm{pb^{-1}}$ of transversely polarized $pp$ data at $\sqrt{s}$ = 200 GeV, respectively. These datasets provide the most precise measurement of the Collins effect in 200 GeV $pp$ collisions to date, especially at the quark momentum fractions $0.1 \le x \le 0.4$. These proceedings present the preliminary results for Collins asymmetries of identified pions in jets in $pp$ collisions at $\sqrt{s}$ = 200 GeV and comparisons to theory predictions.}

\kword{TMD, Collins effect, STAR}

\begin{document}
\maketitle

\section{Introduction}
The transverse spin phenomena in hadron-hadron collisions has gathered worldwide interest in the last few decades. Significant progress has been made to map out the three dimensional tomographic structure of the nucleons through the study of the transverse momentum dependent (TMD) approaches and twist-3 formalism. These studies offer a unique opportunity to explore the correlations between the transverse spin of a nucleon and transverse momenta of the partons inside the nucleon, to test advanced concepts of the factorization, and to investigate the universality and gauge invariance of the quantum chromodynamics (QCD).\par

\begin{figure}[!hbt]
\centering
    \includegraphics[width=0.5\columnwidth]{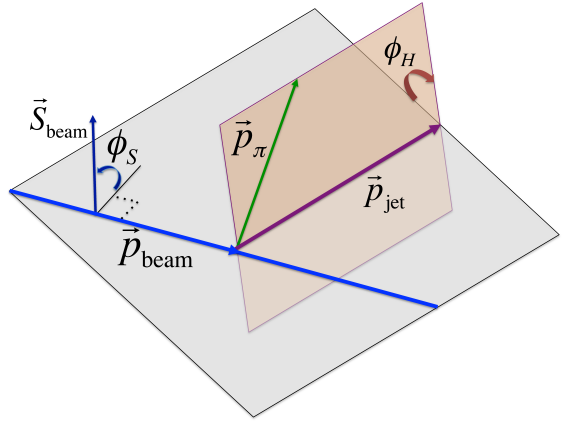}
    \caption{Definition of azimuthal angles $\phi_{S}$ and $\phi_{H}$ in polarized hadronic collisions.}
    \label{fig:azimuthal_modulation}
\end{figure}

The Collins effect~\cite{Collins:1993} is one of the hot topics in TMD physics. It involves the correlation of transverse spin of a quark and the momentum of a hadron fragment transverse to the scattered quark direction. In transversely polarized proton-proton collisions, the Collins asymmetry for charge pions inside jets is generated through the correlation of the transverse spin of the fragmenting quark with the transverse momentum of the hadron with respect to the jet axis. Following the same definition as in Ref.~\cite{Alesio:2011PRD, Alesio:2017PLB}, as shown in Fig.~\ref{fig:azimuthal_modulation}, we can define $\phi_{S}$ as the azimuthal angle between the polarization of the proton beam to the jet scattering plane formed by the jet momentum and beamline in the lab frame, and $\phi_{H}$ as the azimuthal angle of the hadrons inside the jet relative to the jet scattering plane. For hadrons within jets, the spin dependent cross section are the combinations of the azimuthal modulations with different differential cross section terms as shown in Eq.~\ref{equ:spin_cross_section}~\cite{Alesio:2011PRD, Alesio:2017PLB}:\par
\begin{equation}
\label{equ:spin_cross_section}
\begin{split}
&d\sigma^{\uparrow}(\phi_{S}, \phi_{H}) - d\sigma^{\downarrow}(\phi_{S}, \phi_{H}) \\
&\sim d\Delta \sigma_{1}^{-} \sin(\phi_{S} - \phi_{H}) + d\Delta \sigma_{1}^{+} \sin(\phi_{S} + \phi_{H})\\
&+    d\Delta \sigma_{2}^{-} \sin(\phi_{S} - 2\phi_{H}) + d\Delta \sigma_{2}^{+} \sin(\phi_{S} + 2\phi_{H})\\
\end{split}
\end{equation}
where the $d\Delta \sigma$ terms represent various combinations of the TMD parton distribution functions and fragmentation functions, as well as the hard partonic scattering amplitudes that can be calculated from perturbative QCD. The TMD parton distribution functions contain all the information about the polarization state of the initial parton, which depends on the soft, nonperturbative dynamics encoded in the eight leading-twist polarized and transverse momentum dependent parton distribution functions. And the fragmentation functions describe the fragmentation process of the scattered (polarized) parton into charged pions inside the jet.\par

The transverse single spin asymmetry with the modulation of $\sin{(\phi_{S} - \phi_{H})}$ can be defined in terms of the spin dependent cross section and expressed as in Eq.~\ref{equ:spin_cross_section_AN}:
\begin{equation}
\label{equ:spin_cross_section_AN}
A_{N}\sin(\phi) = \frac{\sigma^{\uparrow}(\phi) - \sigma^{\downarrow}(\phi)}{\sigma^{\uparrow}(\phi) + \sigma^{\downarrow}(\phi)} 
\rightarrow
\frac{\sum_{abc}h_{1}^{a}(x_{1},\mu)f_{b}(x_{2},\mu)\sigma^{\mathrm{Collins}}_{ab\rightarrow c}H_{1,h/c}^{\perp}(z_{h},j_{T};Q)}{\sum_{abc}f_{a}(x_{1},\mu)f_{b}(x_{2},\mu)\sigma^{\mathrm{unpol}}_{ab\rightarrow c}D_{h/c}(z_{h},j_{T};Q)}
\end{equation}
where $h_{1}^{a}(x_{1},\mu)$ is the quark collinear transversity, while $H_{1,h/c}^{\perp}(z_{h},j_{T};Q)$ is the TMD Collins fragmentation function. $\sigma^{\mathrm{unpol}}$ is the unpolarized partonic cross section while $\sigma^{\mathrm{Collins}}$ is the spin-dependent partonic cross section. Collins asymmetry in $pp$ collisions involves the collinear transversity with the TMD Collins fragmentation function. The collinear transversity, $h^{a}_{1}(x,\mu)$, depends only on the longitudinal momentum fraction ($x$) and factorization scale ($\mu$), while the TMD Collins fragmentation function, $H_{1,h/c}^{\perp}(z_{h},j_{T};Q)$, depends on the momentum fraction of the fragmenting quark carried by the hadron ($z_{h}$), the hadron transverse momentum with respect to the jet axis ($j_{T}$) and the TMD evolution scale ($Q$). This separation and independence of the TMD parton distribution functions (PDFs) allows a direct probe of TMD fragmentation functions (FFs)~\cite{Kang:2017JHEP, Kang:2017PLB}.\par

The Relativistic Heavy Ion Collider (RHIC)~\cite{Alekseev:2003sk} at Brookhaven National Laboratory provides a unique opportunity to explore the transverse spin phenomena, through collisions of polarized proton beams at center-of-mass energies $\sqrt{s}=$ 200 and 510 GeV. In 2018, STAR published the first measurements of Collins asymmetries for jet + $\pi^{\pm}$ production in polarized $pp$ collisions at $\sqrt{s}=$ 500 GeV~\cite{STARjet2011pp500} based on the 25 $\mathrm{pb^{-1}}$ data sample taken during 2011. In 2012 and 2015, STAR recorded datasets of 14 $\mathrm{pb^{-1}}$ and 52 $\mathrm{pb^{-1}}$, respectively, at $\sqrt{s}=$ 200 GeV, with an average polarization of about 57\%. These results probe higher momentum scales ($Q^{2}$  $\sim$ 960 $\mathrm{GeV^{2}}$ for 500 GeV and $\sim$ 170 $\mathrm{GeV^{2}}$ for 200 GeV) than the measurements from SIDIS ($Q^{2}$  $<$ 20 $\mathrm{GeV^{2}}$)~\cite{HERMES2009, HERMES2010, HERMES2020, COMPASS2009, COMPASS2015, COMPASS2017, JLabHallA2011, JLabHallA2014}, and enable tests of evolution, universality and factorization breaking in the TMD formalism.\par

\section{Jet Reconstruction and Particle Identification}
Jets are reconstructed using the anti-$k_{T}$ algorithm \cite{antikt} with the radius $R$ = 0.6 for the 200 GeV measurement~\cite{STARjet2009pp200}. Charged particles measured by the STAR Time Projection Chamber (TPC)~\cite{Anderson:2003ur} and energy deposits in the Barrel and Endcap Electromagnetic Calorimeter (BEMC and EEMC)~\cite{Beddo:2002zx, Allgower:2002zy} are taken as inputs into the fastjet package~\cite{fastjet}. To avoid double counting of charged particle energy in the EMC, momentum of a charged particle is subtracted from the EMC tower energy if it matches to the tower.\par

The off-axis cone method~\cite{STARjetdijet2012pp500} is adopted to correct for the underlying event contribution in the analysis. In this method, two off-axis cones, with the same radius as jets and located at the same jet $\eta$ but $\pm \pi/2$ away in $\phi$, are identified. The average activity inside these two cones is used as an estimate of underlying event inside the jets. Both the jet energies and spin asymmetries are corrected for the smearing from the underlying event contamination.\par

\begin{figure}[!hbt]
\centering
    \includegraphics[width=0.5\columnwidth]{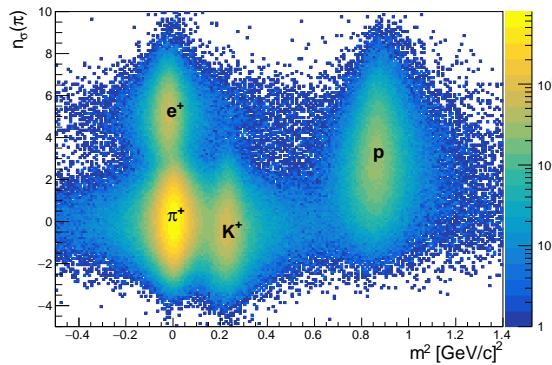}
    \caption{Correlations of $n_\sigma(\pi)$ vs. $m^{2}$ for positively charged particles carrying momentum fractions of 0.1 $< z <$ 0.13 in jets with 8.4 $< p_{T} <$ 9.9 GeV$/c$.}
    \label{fig:dEdx_tofm2_2D}
\end{figure}

Charged hadrons are selected if the measured energy loss ($n_\sigma(\pi)$) in the TPC is consistent with the expected values for pions in order to limit the contamination from other types of particles. To improve the particle identification, the Time of Flight (TOF)~\cite{STARTOF2005} detector is used when $n_\sigma(\pi)$ of two different particles are close. The mass square of a particle can be calculated by $m^{2} = p^{2}(1/\beta^{2}-1)$ with the momentum ($p$) measured from TPC and the inverse velocity $1/\beta$ from TOF. As can be seen from Fig.~\ref{fig:dEdx_tofm2_2D}, TOF provides very good separation of different particle species when their energy losses are close~\cite{STARPID2006, STARPID2010}.\par

\section{Results}
\begin{figure*}
\centering
\begin{minipage}{0.48\columnwidth}
  \centering
 \includegraphics[width=0.9\linewidth]{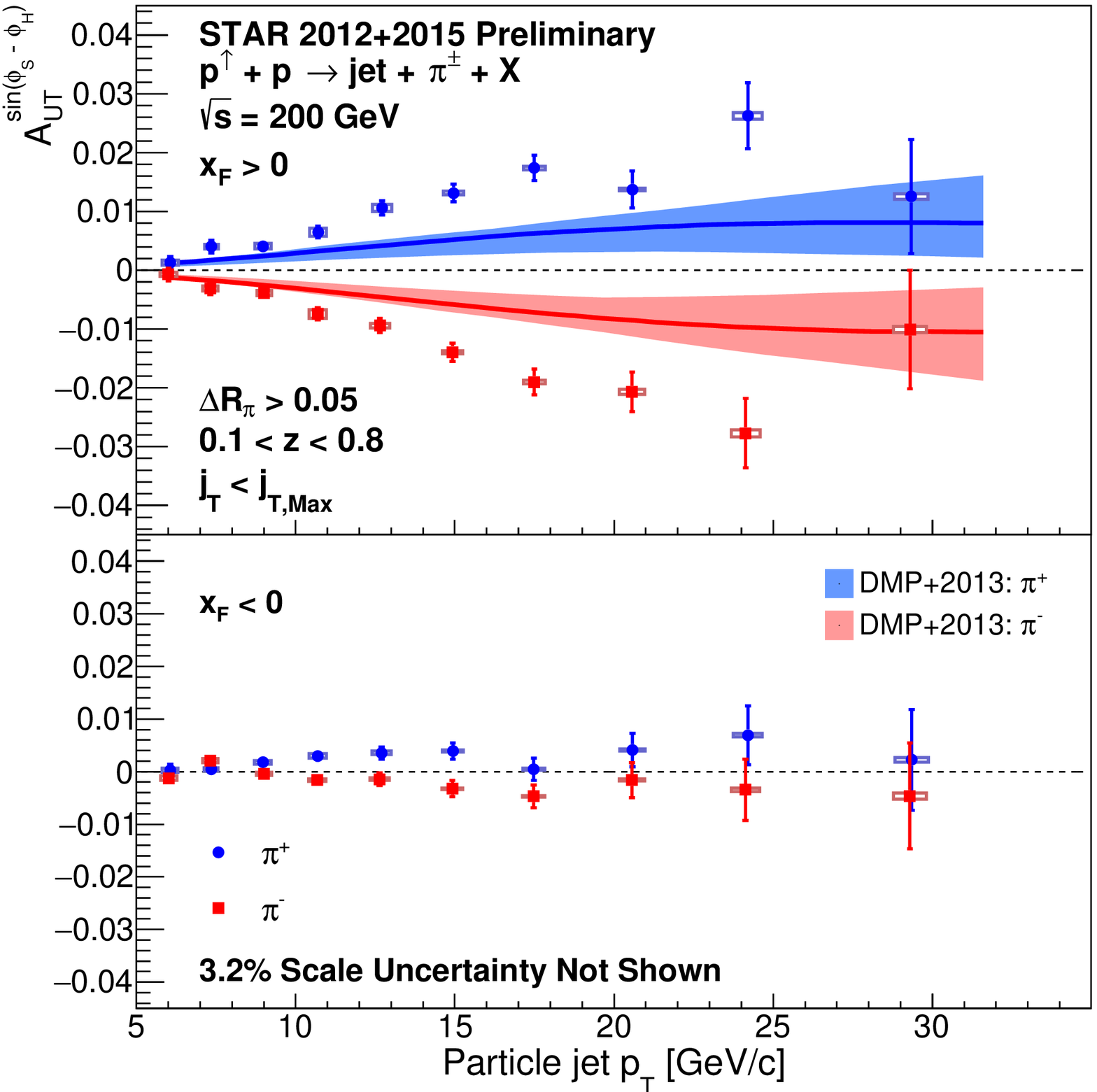}
  \caption{Collins asymmetry as a function of jet $p_{T}$. The blue points represent $\pi^{+}$ and the red ones are for $\pi^{-}$. Top panel is for $x_{F} > 0$ while bottom panel is for $x_{F} < 0$.}
  \label{fig:figure_collins_jet_pt}
\end{minipage}%
\hfill
\begin{minipage}{0.48\columnwidth}
  \centering
  \includegraphics[width=0.9\linewidth]{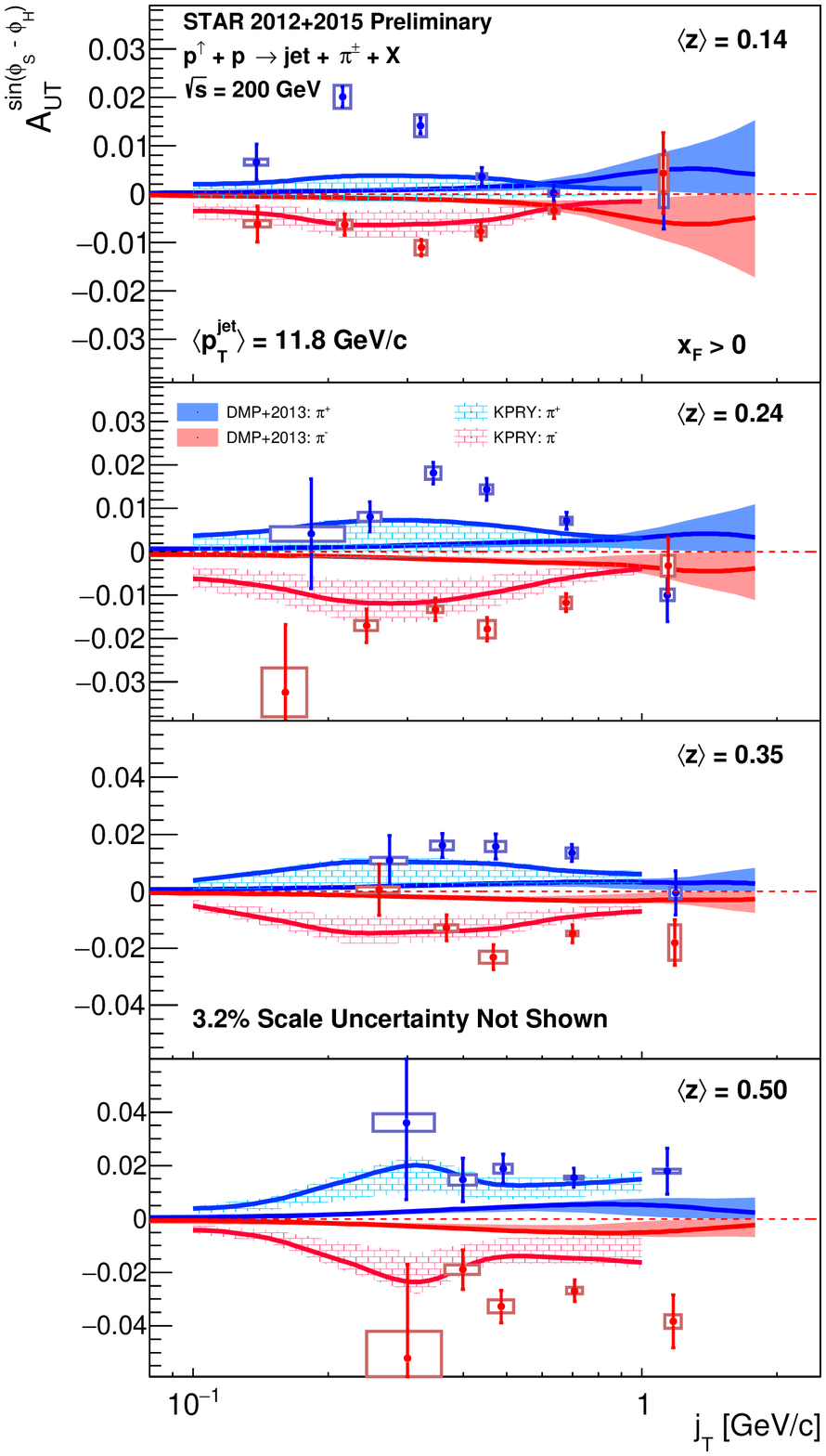}
  \caption{Collins asymmetry as a function of pion momentum transverse to the jet thrust axis, $j_{T}$, in four different pion longitudinal momentum fraction bins. The blue points represent $\pi^{+}$ and the red ones are for $\pi^{-}$. Results shown here are for $x_{F} > 0$.}
  \label{fig:figure2_jT}
\end{minipage}
\end{figure*}

The preliminary results of the Collins asymmetries for charged pions within jets in $pp$ collisions at $\sqrt{s}$ = 200 GeV are presented in Fig.~\ref{fig:figure_collins_jet_pt} as a function of the jet transverse momentum ($p_{T}$), and in Fig.~\ref{fig:figure2_jT} as a function of the hadron $j_{T}$ in four different $z$ bins. In both figures, blue circles are for $\pi^{+}$ while red squares are for $\pi^{-}$. In Fig.~\ref{fig:figure_collins_jet_pt}, results are divided into two different pseudorapidity ranges. Top panel presents asymmetries for jets that are scattered forward ($x_{F} > 0$) with respect to the polarized beam while the bottom panel shows jets scattered backward ($x_{F} < 0$) with respect to the polarized beam. In Fig.~\ref{fig:figure2_jT}, only the results with $x_{F} > 0$ are presented.\par

There are also theoretical calculations shown in the figures with the same color scheme as data for $x_{F} > 0$. The solid lines are the central values with the uncertainties represented as filled bands. DMP+2013 model~\cite{Alesio:2017PLB, PhysRevD.92.114023} is based on the transversity and Collins fragmentation function from SIDIS~\cite{HERMES2009, HERMES2010, HERMES2020, COMPASS2009, COMPASS2015, COMPASS2017, JLabHallA2011, JLabHallA2014} and $e^{+}e^{-}$ processes~\cite{BELLE2008, BaBar2014, BaBar2015, BESIII2016} with leading order TMD approach. KPRY model~\cite{Kang:2017PLB} is also based on the global analyses of SIDIS and $e^{+}e^{-}$ processes~\cite{PhysRevD.93.014009} with TMD evolution up to the next-to-leading-logarithmic order. As can be seen from Fig.~\ref{fig:figure2_jT}, the peak positions of the measured asymmetries are $j_{T}$ and $z$ dependent, which are not observed in any of the models. And in both figures, measured asymmetries are larger than the theoretical calculations, which may indicate larger transversity than the current expectations in this kinematic region.\par

\section{Conclusion}
In summary, we presented new preliminary results of Collins asymmetries for charged pions inside jets in 200 GeV $pp$ collisions from the STAR experiment. Significant Collins asymmetries have been observed, which provide constraints to both the collinear transversity and TMD Collins fragmentation functions at much higher $Q^{2}$ values than the measurements from SIDIS. The measured asymmetries are larger than the theoretical calculations which may indicate larger quark transversity. There is also an ongoing analysis using 510 GeV $pp$ dataset from 2017 ($\sim$ 350 $\mathrm{pb^{-1}}$, ~13 times more than 2011 data), which will provide precise measurements at a lower momentum fraction region than those at 200 GeV. The STAR forward upgrades provide a unique opportunity to explore the TMD physics at forward rapidity with the polarized $pp$ collisions to be taken in 2022 and 2024.\par

\end{document}